\documentclass{ws-mpla}

\begin{document}

\markboth{WEI CHEN, PU-QING ZHANG, LIANG-GANG LIU} {The Influence
of the Magnetic Field......}

%
\catchline{}{}{}{}{}
%

\title{The Influence of the Magnetic Field on the Properties of
Neutron Star Matter}

\author{\footnotesize WEI CHEN\footnote{
Huangpu Road West 601, Guangzhou}}

\address{Department of Physics, Jinan University, Huangpu Road West 601\\
Guangzhou, 510632, China\\
tchenw@jnu.edu.cn}

\author{PU-QING ZHANG}

\address{Department of Physics, Jinan University, Huangpu Road West 601\\
Guangzhou, 510632, China}
\author{LIANG-GANG LIU}
\address{Department of Physics, Zhongshan University, Xingang West Road 135\\
Guangzhou, 510275, China}
\maketitle

\pub{Received (Day Month Year)}{Revised (Day Month Year)}

\begin{abstract}
In the mean field approximation of the relativistic
$\sigma$-$\omega$-$\rho$ model, the magnetic fields are
incorporated, and its influence on the properties of n-p-e neutron
star matter are studied. When the strength of the magnetic field
is weaker than $\sim 10^{18}$G, the particle fractions and
chemical potentials, matter energy density and pressure hardly
change with the magnetic field; when the strength of the magnetic
field is stronger than $\sim 10^{20}$G, the above quantities
change with the magnetic field evidently. Furthermore, the
pressure is studied in both thermodynamics and hydrodynamics. The
difference between these two ways exits in the high density
region, that is, the thermal self-consistency may not be satisfied
in this region if the magnetic field is considered.

\end{abstract}

\ccode{PACS Nos.:97.60.jd, 26.60.+c}

\section{\label{Introduction}Introduction}
Over the past several ten years, the observations have indicated
that large magnetic fields exist at the surface of the neutron
stars\cite{wkvh}. From the spindown rates of pulsars, the magnetic
fields of 558 pulsars of the catalog by Taylor\cite{tml} lies
between $B=1.7\times 10^{8}$G(PSR B1957+20)and $B=2.1\times
10^{13}$G (PSR B0154+61), with a typical value $B=1.3\times
10^{12}$G, most of young pulsars having a surface field in the
range $B\sim 0.1-2\times 10^{13}$G. Several proofs suggest that
soft $\gamma$-ray repeaters(SGRS), and perhaps the so-called
anomalous X-ray pulsars(AXPS), are neutron stars with magnetic
fields $\geq 10^{14}$G-----the so called magnetars\cite{vg}.
Furthermore, the observed X-ray luminosities of the AXPs may
require a field strength $B\geq 10^{16}$G\cite{chn}. The
population statics of SGRs suggest that magnetars may constitute a
significant fraction($\geq 10\%$) of the neutron star
population\cite{kdsvfmh}. A considerable amount of studies have
been devoted to the structure of the magnetic field outside the
neutron star, in the so-called magnetosphere, in relation with the
pulsar emission mechanism(for a review, see e.g.\cite{Michel}).
The strength of the magnetic field in the interior of neutron
stars remains unknown, its estimated values can be $10^{18}$G for
a star with $R\approx 10$km and $M\approx
1.4M_{\bigodot}$\cite{st}.\\
\indent The effect of the neutron star strong interior magnetic
fields has been studied recently\cite{bbgn,cpl}. Broderick etc.
found that the magnetic fields of strengths larger than $10^{16}$G
affect the equation of state(EOS) of dense matter directly through
drastic changes in the composition of matter\cite{bpl}. The EOS is
altered by both the landau quantization of the charged
particles(such as protons, electrons, etc.)and the interactions of
the magnetic moments, including the anomalous magnetic moments of
the neutral particles(such as the neutron, strangness-bearing
$\Lambda$-hyperon etc.), with the
magnetic field.\\
\indent We study the properties of n-p-e neutron star by
incorporating the magnetic field with the relativistic
$\sigma$-$\omega$-$\rho$ model. In other works, the pressure is
usually studied by thermodynamics. As we know, the pressure can be
calculated by both thermodynamics and hydrodynamics, which are
equivalent in nuclear matter when the magnetic field is absent.
Considering the magnetic field in neutron stars, the difference
between these two ways will exist as pointed out in this paper,
especially in the high density region. The theoretical formalism
is outlined in section 2; in section 3, the numerical results are
presented; and finally, we give our summary in section 4.

\section{\label{Basic theory}Basic theory}
The Lagrangian density is given by:
\begin{eqnarray}
{\cal L}&=&\bar{\psi}[i\gamma_{\mu}D^{\mu}-m+g_{s}\phi
-\gamma_{\mu}g_{v}\omega^{\mu}-
\frac{1}{2}g_{\rho}\gamma_{\mu}\vec{\tau}\cdot\vec{\rho^{\mu}}
]\psi+\frac{1}{2}\partial^{\mu}\phi\partial_{\mu}\phi-
\frac{1}{2}m_{s}^{2}\phi^{2}-\nonumber\\
&&\frac{1}{4}\omega_{\mu\nu}\omega^{\mu\nu}-\frac{1}{4}\rho_{\mu\nu}\rho^{\mu\nu}+
\frac{1}{2}m_{\omega}^{2}\omega_{\mu}\omega^{\mu}+
\frac{1}{2}m_{\rho}^{2}\vec{\rho}_{\mu}\cdot\vec{\rho}^{\mu}+
\bar{\psi}_{e}[i\gamma_{\mu}D^{\mu}-m_{e}]\psi_{e}-\nonumber\\
&&U(\sigma) \label{eq.1},
\end{eqnarray}

\begin{equation}
U(\sigma)=\frac{1}{3}c\phi^{3}+\frac{1}{4}d\phi^{4}
\end{equation}

\begin{equation}
\omega_{\mu\nu}=\partial_{\mu}\omega_{\nu}-\partial_{\nu}\omega_{\mu}
\end{equation}

\begin{equation}
\rho_{\mu\nu}=\partial_{\mu}\vec{\rho}_{\nu}-\partial_{\nu}\vec{\rho}_{\mu}
\end{equation}

 $\psi$, $\psi_{e}$, $\phi$, $\omega^{\mu}$ and
$\vec{\rho}^{\mu}$ are the fields of nucleon, electron, $\sigma$,
$\omega$, $\rho$ meson respectively. $g_{s}$, $g_{v}$ and
$g_{\rho}$ are the coupling constants of $\sigma$, $\omega$,
$\rho$ meson to nucleons respectively. $m$, $m_{e}$, $m_{s}$,
$m_{\omega}$ and $m_{\rho}$ are the mass of nucleon, electron,
$\sigma$, $\omega$ and $\rho$ meson respectively.

\begin{eqnarray}
D^{\mu}=\partial^{\mu}+iqA^{\mu}\nonumber\\
A^{0}=0, \vec{A}=(0, xB_{m}, 0)
 \label{eq.2},
\end{eqnarray}

$q$ is the charge of a particle, $A^{\mu}=(A^{0}, \vec{A})$ is the
four dimensional electromagnetic vector, we select it as
Eq.~(\ref{eq.2}) to obtain a magnetic field along $z$ axis. We
don't consider the anomalous magnetic moments, so neutrons are not
affected by magnetic field due to its neutrality. The charge of
proton is $e$, its move equation, the formalism of the field
operator and eigenvalue are different from the case in which the
magnetic field is ignored. And its move equation is:

\begin{equation}
i\gamma_{\mu}\partial^{\mu}\psi_{p}-[m-g_{s}\phi
+\gamma_{\mu}g_{v}\omega^{\mu}+
\frac{1}{2}g_{\rho}\gamma_{\mu}\vec{\tau}\cdot\vec{\rho^{\mu}}
+e\gamma_{\mu}A^{\mu}]\psi_{p}=0 \label{eq.3},
\end{equation}
the general solution of Eq.~(\ref{eq.3}) is:

\begin{equation}
\psi_{p}\approx
\exp^{-i\varepsilon^{H}t+ip_{y}y+ip_{z}z}f_{p_{y},p_{z}}(x)
\label{eq.4},
\end{equation}
where $f_{p_{y},p_{z}(x)}$ is the four-component solution, and
Eq.~(\ref{eq.3}) can be converted into:

\begin{equation}
[-i\alpha_{x}\frac{\partial}{\partial
x}+\alpha_{y}(p_{y}-exB_{m})+\beta
m^{*}+\alpha_{z}p_{z}+U_{0,p}^{H}]f_{p_{y},p_{z}}(x)=\varepsilon^{H}f_{p_{y},p_{z}}(x)
\label{eq.5},
\end{equation}

\begin{equation}
m^{*}=m-g_{s}\upsilon \label{eq.6},
\end{equation}
$\upsilon$ is the mean field of $\sigma$ meson field,
\begin{equation}
U_{0,N}^{H}=\frac{g_{\omega}^{2}}{m_{\omega}^{2}}\rho_{B}+
\frac{1}{4}\frac{g_{\rho}^{2}}{m_{\rho}^{2}}\rho_{3}I_{3N},
\,\,\,\,\,\,N=p,\,\,\,n \label{eq.7},
\end{equation}
$I_{3N}$ is the projection of nucleon isospin along $z$ axis,
$I_{3p}=1$, $I_{3n}=-1$.

\begin{equation}
\rho_{B}=\rho_{p}+\rho_{n},\,\,\,\,\, \rho_{3}=\rho_{p}-\rho_{n}
\label{eq.8},
\end{equation}
$\rho_{B}$, $\rho_{n}$, $\rho_{p}$ and the following $\rho_{e}$
are the density of baryons, neutrons, protons and electrons
respectively. By resolving the Eq.~(\ref{eq.5}), $\psi_{p}$ is
obtained:
\begin{equation}
\psi_{p1}^{(+)}(x)=
\frac{\exp^{(-i\varepsilon^{H}t+ip_{y}y+ip_{z}z)}}{\sqrt{2\varepsilon_{\nu}^{H}(\varepsilon_{\nu}^{H}+p_{z})}}
\left(\begin{array}{c}
(\varepsilon_{\nu}^{H}+p_{z})I_{\nu,p_{y}}(x)\\
-i\sqrt{2eB_{m}\nu}I_{\nu-1,p_{y}}(x)\\
-m^{*}I_{\nu,p_{y}}(x)\\
0 \end{array} \right)
 \label{eq.9},
\end{equation}

\begin{equation}
\psi_{p1}^{(-)}(x)=
\frac{\exp^{(i\varepsilon^{H}t-ip_{y}y-ip_{z}z)}}{\sqrt{2\varepsilon_{\nu}^{H}(\varepsilon_{\nu}^{H}-p_{z})}}
\left(\begin{array}{c}
-m^{*}I_{\nu,-p_{y}}(x)\\
0\\
(-\varepsilon_{\nu}^{H}+p_{z})I_{\nu,-p_{y}}(x)\\
i\sqrt{2eB_{m}\nu}I_{\nu-1,-p_{y}}(x) \end{array} \right)
 \label{eq.10},
\end{equation}

\begin{equation}
\psi_{p2}^{(+)}(x)=
\frac{\exp^{(-i\varepsilon^{H}t+ip_{y}y+ip_{z}z)}}{\sqrt{2\varepsilon_{\nu}^{H}(\varepsilon_{\nu}^{H}+p_{z})}}
\left(\begin{array}{c}
0\\
-m^{*}I_{\nu-1,p_{y}}(x)\\
-i\sqrt{2eB_{m}\nu}I_{\nu,p_{y}}(x)\\
(\varepsilon_{\nu}^{H}+p_{z})I_{\nu-1,p_{y}}(x) \end{array}
\right)
 \label{eq.11},
\end{equation}

\begin{equation}
\psi_{p2}^{(-)}(x)=
\frac{\exp^{(i\varepsilon^{H}t-ip_{y}y-ip_{z}z)}}{\sqrt{2\varepsilon_{\nu}^{H}(\varepsilon_{\nu}^{H}-p_{z})}}
\left(\begin{array}{c}
i\sqrt{2eB_{m}\nu}I_{\nu,-p_{y}}(x) \\
(-\varepsilon_{\nu}^{H}+p_{z})I_{\nu-1,-p_{y}}(x)\\
0\\
-m^{*}I_{\nu-1,-p_{y}}(x)\end{array} \right)
 \label{eq.12},
\end{equation}

\begin{equation}
\varepsilon_{\nu}^{H}=\sqrt{p_{z}^{2}+m^{*2}+2eB_{m}\nu}=\varepsilon^{H}-U_{0,p}^{H}
\label{eq.13},
\end{equation}

$I_{\nu,p_{y}}(x)$ is normalized as:
\begin{equation}
\int dxI_{\nu,p_{y}}(x)I_{\nu^{'},p_{y}}(x)=\delta_{\nu\nu^{'}},
\,\,\,\,\,\,\,\,\,
\sum_{n=0}^{\infty}I_{\nu,p_{y}}(x)I_{\nu,p_{y}}(x^{'})=\delta
(x-x^{'}) \label{eq.14}.
\end{equation}
$\nu$ is the Landau quantum number, $\psi_{e}$ has the same form
as $\psi_{p}$ with the proton quantities replaced by the electron
ones. The proton propagator can be constructed by $\psi_{p}$.
Furthermore, by the known neutron propagator, we can present the
energy density $\varepsilon$ and pressure $p$ in the mean field
approximation:

\begin{eqnarray}
\varepsilon=&&\frac{\gamma_{n}}{(2\pi)^{3}}\int_{k_{F}}d^{3}k\sqrt{k^{2}+m^{*2}}+
U_{0,n}^{H}\rho_{n}+\frac{eB_{m}}{(2\pi)^{2}}\sum_{\nu=0}^{\nu_{max}}g_{\nu}
\int_{-p_{3F}}^{p_{3F}}dp_{z}\varepsilon_{\nu}^{H}+\nonumber\\&&U_{0,p}^{H}\rho_{p}+
\frac{eB_{m}}{(2\pi)^{2}}\sum_{\nu=0}^{\nu_{max}^{(e)}}g_{\nu}
\int_{-p_{3Fe}}^{p_{3Fe}}dp_{ze}\varepsilon_{\nu e}^{H}+
\frac{1}{2}m_{s}^{2}\upsilon^{2}-\frac{1}{2}m_{\omega}^{2}V_{0}^{\omega
2}-\nonumber\\&&\frac{1}{2}m_{\rho}^{2}V_{0}^{\rho
2}+\frac{1}{3}c\upsilon^{3}+\frac{1}{4}d\upsilon^{4}
\label{eq.15},
\end{eqnarray}

\begin{eqnarray}
p=&&\frac{1}{3}\frac{\gamma_{n}}{(2\pi)^{3}}\int_{k_{F}}d^{3}k\frac{k^{2}}{\sqrt{k^{2}+m^{*2}}}+
\frac{1}{3}\frac{eB_{m}}{(2\pi)^{2}}\int_{-p_{3F}}^{p_{3F}}dp_{z}\sum_{\nu=0}^{\nu_{max}}g_{\nu}
\frac{p_{z}^{2}+\nu eB_{m}}{\varepsilon_{\nu}^{H}}+\nonumber\\
&&\frac{1}{3}\frac{eB_{m}}{(2\pi)^{2}}\int_{-p_{3Fe}}^{p_{3Fe}}dp_{ze}\sum_{\nu=0}^{\nu_{max}^{(e)}}
g_{\nu}\frac{p_{ze}^{2}+\nu eB_{m}}{\varepsilon_{\nu e}^{H}}-
\frac{1}{2}m_{s}^{2}\upsilon^{2}+\frac{1}{2}m_{\omega}^{2}V_{0}^{\omega
2}+\nonumber\\
&&\frac{1}{2}m_{\rho}^{2}V_{0}^{\rho
2}-\frac{1}{3}c\upsilon^{3}-\frac{1}{4}d\upsilon^{4}
\label{eq.16},
\end{eqnarray}

\begin{equation}
\rho_{p}=\frac{eB_{m}}{2\pi^{2}}\sum_{\nu=0}^{\nu_{max}}p_{3F}g_{\nu}
\label{eq.17},
\end{equation}

\begin{equation}
\rho_{e}=\frac{eB_{m}}{2\pi^{2}}\sum_{\nu=0}^{\nu_{max}^{(e)}}p_{3Fe}g_{\nu}
\label{eq.18},
\end{equation}

\begin{equation}
\rho_{n}=\frac{\gamma_{n}}{(2\pi)^{3}}\int_{k_{F}}d^{3}k
\label{eq.19},
\end{equation}
chemical equilibrium and neutrality conditions in neutron stars
give
\begin{equation}
\mu_{n}=\mu_{p}+\mu_{e}, \,\,\,\,\,\,\,\,
\rho_{p}=\rho_{e}\label{eq.20}.
\end{equation}
The Landau level degeneracy factor $g_{\nu}$ is 1 for $\nu=0$ and
2 for $\nu>0$. $\mu_{n}$, $\mu_{p}$ and $\mu_{e}$ are the neutron,
proton and electron chemical potential, $V_{0}^{\omega}$,
$V_{0}^{\rho}$ are the mean fields of the time-like component of
the $\omega$ meson field and the time-like isospin 3-component of
$\rho$ meson field respectively.

\section{\label{Numerical calculation}Numerical calculation}
The values of the constants are adjusted to reproduce the

\begin{figure}[th]
\centerline{\psfig{file=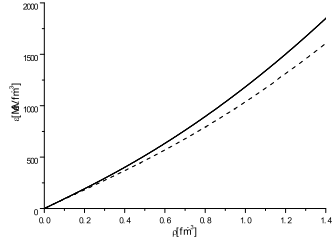,width=2.0in}} \vspace*{8pt}
\caption{The curve of energy density of n-p-e neutron star matter
changing with baryon density in different magnetic field. The
solid and dashed represent the cases in $B_{m}\leq 10^{18}$G,
$B_{m}=10^{20}$G respectively.}\label{fig.1}
\end{figure}

saturation properties of nuclear matter in the mean field
approximation of the relativistic $\sigma$-$\omega$-$\rho$ model,
$g_{s}=7.17$, $g_{\omega}=7.16$, $g_{\rho}=8.55$, $c=100.02$MeV,
$d=372.26$. Fig.~\ref{fig.1} depicts the relation between the
energy density of n-p-e neutron star matter and baryon density in
different magnitude of uniform magnetic field. With the magnetic
\begin{figure}[th]
\centerline{\psfig{file=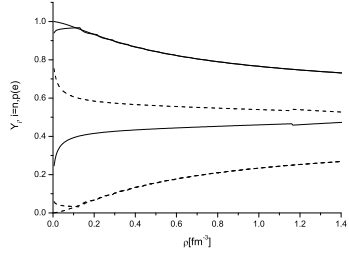,width=2.0in}} \vspace*{8pt}
\caption{The fractions of neutrons(solid) and protons(dashed) of
n-p-e neutron star matter change with baryon density in the
magnetic field of $\leq 10^{16}$G, $10^{18}$G, $10^{20}$(neutron
is from top to bottom and proton is adverse) respectively.}
\label{fig.2}
\end{figure}
field increasing, the energy density doesn't change if $B_{m}\leq
10^{18}$G and decreases rapidly if $B_{m}$ accesses to $10^{20}$G.
In Fig.~\ref{fig.2} we plot the particle fraction,
$Y_{i}=\frac{\rho_{i}}{\rho_{B}}$, $i=n$, $p$, as a function of

\begin{figure}[th]
\centerline{\psfig{file=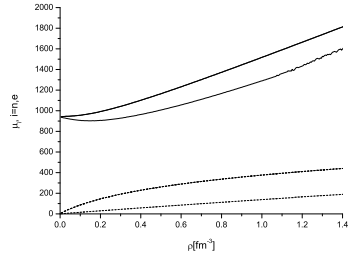,width=2.0in}} \vspace*{8pt}
\caption{The relation between the potential of neutrons(solid),
electrons(dashed) and baryon density in the magnetic field of
$<10^{18}$G, $10^{20}$G(from top to bottom)respectively.}
\label{fig.3}
\end{figure}

the baryon density in the same case as Fig.~\ref{fig.1}. The cases
$B_{m}\leq10^{18}$G are nearly indistinguishable, except for the
low density region. When $B_{m}$ increases to $10^{20}$G, the
neutron fraction decreases and the proton increases drastically.
Fig.~\ref{fig.3} represents the dependence of the particles
chemical potentials on the baryon density in the magnetic field.
Their behavior with the magnetic field is similar to
the one of the particle fractions as in Fig.~\ref{fig.2}. \\
\begin{figure}[th]
\centerline{\psfig{file=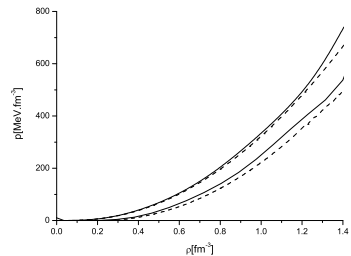,width=2.0in}} \vspace*{8pt}
\caption{The pressure of n-p-e neutron star matter changes with
baryon density in the magnetic field of $<10^{18}$G,
$10^{20}$G(from top to bottom). The solid and the dashed represent
the results by thermodynamics and hydrodynamics respectively.}
\label{fig.4}
\end{figure}
\indent We study the pressure by Eq.(19) as the dashed in
Fig.~\ref{fig.4}. For comparison, we also study it by
thermodynamics, that is
$p=\rho_{B}^{2}\frac{\partial}{\partial\rho_{B}}(\frac{\varepsilon}{\rho_{B}})$.
These two results ,which are equivalent in nuclear matter, are
different in the high density region, and more evident if $B_{m}>
10^{18}$G. Its cause may be that more complex factors are not
considered, for example, the particle magnetic momentum and
anomalous magnetic momentum, the non uniformity of the magnetic
field etc..

\section{\label{Summary}Summary}
\indent By appropriating the magnetic field in the relativistic
$\sigma$-$\omega$-$\rho$ model, we have studied the properties of
n-p-e neutron star matter. It is found that the properties are
nearly invariant with the magnetic field when $B_{m}\leq
10^{18}$G. To observe the effect of the magnetic field clearly, we
assume the value of $B_{m}$ to be $10^{20}$G, regardless of its
physical meaning. As we expected, the properties are altered
evidently. In addition, the discrepancy exits between the
pressures studied by two ways as mentioned above in the high
density region. For simplicity, some complex factors which are
presumed to be present in the interior of neutron star are
omitted. This rude treatment may be just the cause of the
discrepancy between pressures by thermodynamics and
hydrodynamics.\\
\indent Further, the quark phase has been studied at finite
temperature or finite chemical
potential\cite{xql,Chen:2004tb,Gregory:1999pm} in detail. It will
be also valuable that the magnetic field in neutron stars is
studied considering these factors.
\section*{Acknowledgments}
This work is partially supported by the National Natural Science
Foundation with Grant NO. 10275099, 10175096, 10347124  and  the
Natural Science Foundation of Jinan University with Grant No.
640567.

\end{document}